\documentclass[12pt]{iopart}

\usepackage{graphicx}
\begin{document}
\title[Low-Energy Electronic Structure of the  high-$T_c$
Cuprates La$_{2-x}$Sr$_x$CuO$_4$ Studied by ...]{Low-Energy
Electronic Structure of the High-$T_c$ Cuprates
La$_{2-x}$Sr$_x$CuO$_4$ Studied by Angle-resolved Photoemission
Spectroscopy}

\author{T. Yoshida$^1$, X. J. Zhou$^2$, D. H. Lu$^2$, Seiki Komiya$^3$, Yoichi Ando$^3$, H. Eisaki$^4$, T. Kakeshita$^5$, S. Uchida$^6$, Z. Hussain$^7$, Z.-X. Shen$^2$ and A. Fujimori$^1$}

\address{$^1$Department of Complexity Science and Engineering, University of Tokyo, Kashiwa, Chiba
277-8561, Japan}
\address{$^2$Department of Applied Physics and Stanford Synchrotron Radiation Laboratory, Stanford University,
Stanford, CA94305, U. S. A.}
\address{$^3$Central Research Institute of Electric Power Industry, Komae, Tokyo 201-8511,
Japan}
\address{$^4$National Institute of Advanced Industrial Science and Technology, Tsukuba 305-8568,
Japan}
\address{$^5$Superconductivity Research Laboratory, ISTEC, Shinonome, Koto-ku, Tokyo 135-0062,
Japan}
\address{$^6$Department of Physics, University of Tokyo, Bunkyo-ku, Tokyo 113-0033, Japan}
\address{$^7$Advanced Light Source, Lawrence Berkeley National
Lab, Berkeley, CA 94720, U. S. A.}
\ead{teppei@k.u-tokyo.ac.jp}

\begin{abstract}
We have performed a systematic angle-resolved photoemission
spectroscopy (ARPES) study of the high-$T_c$ cuprates
La$_{2-x}$Sr$_x$CuO$_4$, ranging from the underdoped insulator to
the superconductor to the overdoped metal. We have revealed a
systematic doping evolution of the band dispersions and
(underlying) Fermi surfaces, pseudogap and quasi-particle features
under the influence of strong electron-electron interaction and
electron-phonon interaction. The unusual transport and
thermodynamic properties are explained by taking into account the
pseudogap opening and the Fermi arc formation, due to which the
carrier number decreases as the doped hole concentration
decreases.
\end{abstract}

\maketitle

\section{Introduction}
The mechanism that causes the high-temperature superconductivity
in the copper oxide materials (cuprates) is still unknown. As the
charge carriers (electrons or holes) are introduced into the
parent antiferromagnetic insulator, the material evolves from an
insulator to a superconductor, and eventually to a normal metal.
Thus, the cuprate systems drastically change their behaviors
depending on the carrier density in the two-dimensional CuO$_2$
planes, which are the stage of the high-$T_{\mathrm c}$
superconductivity and related low-energy physics. While the
overdoped ($x>0.2$) cuprates behave like a relatively conventional
Fermi liquid above $T_{\mathrm c}$, underdoped ($x<0.15$) cuprates
in the normal state show behaviors which are strongly deviated
from those of a standard Fermi liquid. Therefore, it is necessary
to know the detailed information about the doping evolution of the
electronic structure of the CuO$_2$ plane to understand the
cuprate systems.

The recent remarkable development in the angle-resolved
photoemission spectroscopy (ARPES) technique has enabled us to
investigate fine spectral features near the chemical potential,
which may be used to interpret the thermodynamic and transport
properties of solids from a microscopic point of view. Indeed,
ARPES studies on cuprate superconductors, largely on the
Bi$_2$Sr$_2$CaCu$_2$O$_{8+y}$ (Bi2212) family, have been performed
extensively \cite{DamascelliReview} and revealed the normal-state
electronic structure such as the dispersions of quasi-particle
bands crossing the chemical potential, the presence of a
pseudo-gap above $T_c$ and its $d$-wave like gap anisotropy in the
underdoped region. On the other hand, bilayer splitting
\cite{fengB}, BiO-layer structural modulation, etc., complicate
the interpretation of the ARPES spectra of Bi2212.

In order to avoid the above complications and to study the
intrinsic doping dependence of the electronic structure over a
much wider hole concentration range, the La$_{2-x}$Sr$_x$CuO$_4$
(LSCO) system is the most suitable one among the families of
high-$T_{\mathrm c}$ cuprate superconductors. LSCO has a simple
crystal structure with single CuO$_2$ layers, and no Cu-O chains
as in YBa$_2$Cu$_3$O$_{7-y}$ (YBCO) nor complicated structural
modulation of the block layers as in Bi2212. The hole
concentration in the CuO$_2$ plane can be controlled over a wide
range and uniquely determined by the Sr concentration $x$ (and the
small oxygen non-stoichiometry). One can therefore investigate the
doping dependence from the undoped insulator ($x = 0$) to the
heavily overdoped metal ($x=0.3$) in a single system. Indeed, the
most extensive set of data on the doping dependence of the
transport and thermodynamic properties are available for the LSCO
system
\cite{nakano1,hwang,nishikawa1,takagi1,loram1,nishikawa2,momono}
and can be in principle compared with ARPES data.

ARPES studies of LSCO had been behind those of Bi2212 for some
time due to the experimental difficulties in obtaining atomically
flat surfaces by cleaving and to the relatively poor surface
chemical stability. The first successful ARPES study of LSCO was
performed by Ino \textit{et al.} \cite{ino1}. They showed an
overall feature of band dispersions and Fermi surface, including a
crossover from a hole-like to an electron-like Fermi surface with
hole doping \cite{ino1,ino2,ino3}. Peculiar features such as an
unusual suppression of the zone diagonal direction, which had not
been observed in Bi2212, were also reported in \cite{ino1} and
subsequently by Zhou \textit{et al.} \cite{XJZstripePRL}. In
addition, so-called ``two-component" behavior, namely, the
coexisting remnant lower Hubbard band and quasi-particle band with
the chemical potential pinned within the charge-transfer gap, have
been observed in the underdoped region \cite{ino2}. Those
unconventional behaviors were discussed in terms of microscopic
charge inhomogeneity or the spin-charge fluctuations in a stripe
form \cite{zaanen,salkola}, which has been suggested from
incommensurate peaks in the inelastic neutron scattering studies
\cite{tranquada}. Alternatively, the apparent ``two-component"
behaviors in the ARPES line shapes have been interpreted by the
strong electron-phonon coupling \cite{KMShenPRL,Rosch,Nagaosa}.

Because of the weakness of the nodal spectral weight, concerns
about the sample surface quality had persisted. Also, there were
questions about the effects of photoemission matrix elements
\cite{bansil}, which makes it difficult to extract quantitative
information about the stripe effects from comparison between
experiment and theoretical studies which predicted the suppression
of spectral weight in the nodal direction
\cite{tohyama,fleck,ichioka,machida}. However, recent ARPES
studies on LSCO have revealed that under a suitable experimental
geometry, a quasi-particle (QP) band with a clear kink feature
crosses the Fermi level ($E_F$) in the nodal (0,0)-($\pi$,$\pi$)
direction \cite{YoshidaOD} even down to the lightly doped region
\cite{Metallic}. It has also been established that the Fermi
surface is basically large \cite{Luttinger} and that the ($\pi$,0)
region remains (pseudo) gapped, leading to the electronic state in
which only part of the Fermi surface survives as an ``arc" around
the nodal direction. Such a large Fermi surface picture with the
pseudogap has been theoretically studied both from RVB
\cite{fukuyama} and Fermi liquid approach \cite{yanase}. In this
papar, we present a large set of ARPES spectra of LSCO showing the
QP structure or the pseudogap feature on the entire Fermi surface,
and compare the ARPES spectra with the transport and thermodynamic
properties in the entire doping range. Particularly, we
demonstrate that, in the underdoped region, both the ARPES spectra
and the physical properties are highly affected by the opening of
the pseudogap and the concomitant formation of the Fermi arc.

\begin{figure}
\includegraphics[width=6cm]{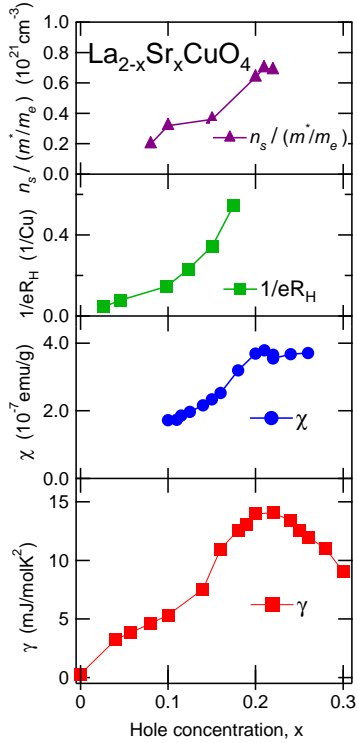}% Here is how to import EPS art
\caption{\label{PG}(Color online) Doping dependence of the
electronic specific coefficient $\gamma$ \cite{momono}, Pauli
paramagnetic suseptivility $\chi$ \cite{nakano1}, the carrier
density deduced from Hall coefficient $R_H$ \cite{takagi}, and the
superfluid density $n_s/m^*$ \cite{uemura} in LSCO.}
\end{figure}

Figure \ref{PG} summarizes the pseudogap behaviors seen in the
thermodynamic and transport properties of LSCO. As shown in the
figure, the electronic specific coefficient $\gamma$ \cite{momono}
and the Pauli paramagnetic suseptibility $\chi$ \cite{nakano1}
decrease with decreasing $x$, in contrast to the mass enhancement
toward the metal-insulator transition usually considered in the
Fermi-liquid picture and experimentally seen in the Fermi-liquid
system La$_{1-x}$Sr$_x$TiO$_3$ \cite{tokuraLSTO}. Furthermore,
although the Fermi surface seen by ARPES is large, the carrier
density $n$ deduced from the Hall coefficient $R_H$ \cite{takagi}
and the superfluid density $n_s$ from magnetic-field penetration
depth \cite{uemura} measurements decrease with decreasing $x$.
That is, the carrier number $n$ appears to be given by the hole
concentration $x$ rather than the band filling $1-x$ in the
underdoped region \cite{takagi,uchida1}. These unconventional
behavior may be attributed to the pseudogap opening through which
a portion of the Fermi surface becomes gapped above $T_c$. The
pseudogap opening provides a natural scenario to the long-standing
issue of how the electronic structure evolves with hole doping
from the Mott insulator to the superconductor in two-dimensional
systems. These features shall be described by the recent
high-quality ARPES data of LSCO as described below.

\section{Experiment}
The ARPES measurements were carried out at beamline 10.0.01 of
Advanced Light Source (ALS) and beamline 5-4 of Stanford
Synchrotron Radiation Laboratory (SSRL), using incident photons
with energies of 55.5 eV and 22.4 eV, respectively. SCIENTA
SES-2002 (ALS) and SES-200 (SSRL) spectrometers were used in the
angle mode. The total energy resolutions were about 15 meV (SSRL)
to 20 meV (ALS). High-quality single crystals of LSCO were grown
by the traveling-solvent floating-zone method. The samples were
cleaved \textit{in situ} and measurements were performed at about
20 K (ALS) or 10 K (SSRL). In the measurements at ALS, the
electric field vector $\mathbf{E}$ of the incident photons lies in
the CuO$_2$ plane, rotated by 45 degrees from the Cu-O bond
direction, so that its direction is parallel to the Fermi surface
segment around the nodal region. This measurement geometry
enhances dipole matrix elements in this $ \mathbf{k}$ region
\cite{YoshidaOD}.

\section{Doping evolution of the electronic structure}

\subsection{Doping Evolution of Quasi-particles}

\begin{figure*}
\includegraphics[width=8cm]{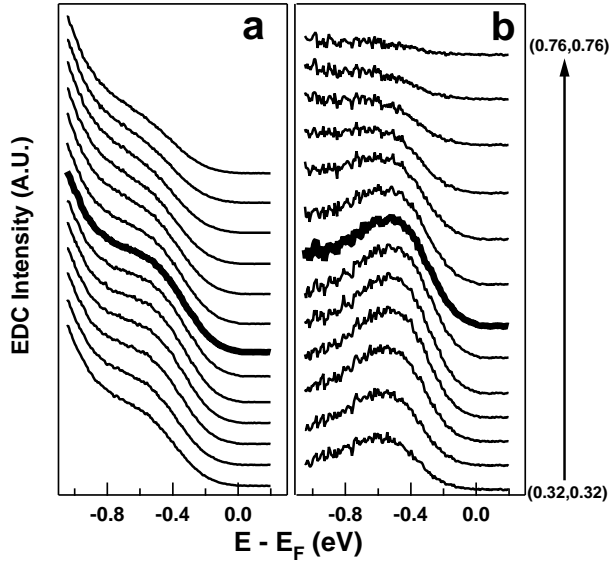}
\caption{\label{EDC0}Photoemission spectra of La$_2$CuO$_4$ along
the (0,0)-($\pi$,$\pi$) nodal direction in the first Brillouin
zone \cite{Rosch}. The corresponding momentum runs from
(0.32,0.32)$\pi$ to (0.76,0.76)$\pi$, as indicated by the arrow.
(a) Raw data. (b) To highlight the momentum dependence, a
"background" given by a spectrum near ($\pi$,$\pi$) has been
subtracted. In both (a) and (b), the bold curves correspond to a
momentum closest to ($\pi$/2,$\pi$/2).}
\end{figure*}

One of the central issues in the field of high $T_c$ cuprates is
how an antiferromagnetic insulator evolves with hole doping.
Figure \ref{EDC0} reveals some important features such as the
doping-dependent dispersions along the (remnant) Fermi surface and
spectral weight transfer from the high-energy Hubbard band to the
low-energy QP band. In order to further look into this problem,
first, we show the energy distribution curves (EDC's) of the
lowest energy feature along the nodal direction of undoped
La$_2$CuO$_4$ . One can see one broad dispersive feature
approximately folded back with respect to the ($\pi/2, \pi/2$)
point \cite{Rosch}, similar to the results for Ca$_2$CuO$_2$Cl$_2$
\cite{KMShenPRL}. As in the case of Sr$_2$CuO$_2$Cl$_2$, the
dispersion of this feature agrees with that of
$t-t^\prime-t^{\prime\prime}-J$ model well
\cite{Wells,Maekawa,Dagotto}. The broad band at $\sim0.5$ eV has
been interpreted as a boson side band showing a Franck-Condon
broadening. A shell-model calculation showed that the width of the
phonon sideband is in good agreement with the ARPES data of
undoped La$_2$CuO$_4$ \cite{Rosch}. This indicates that the
electron-phonon coupling is strong enough, which seems a universal
feature in the undoped cuprates.

\begin{figure}
\includegraphics[width=14.0cm]{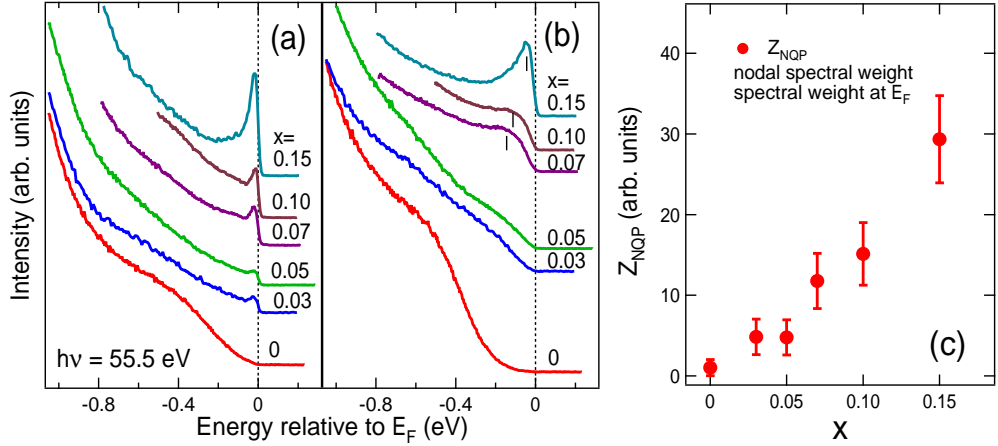}% Here is how to import EPS art
\caption{\label{EDC}(Color online) ARPES spectra at $\mathbf{k} =
\mathbf{k}_\mathrm{F}$ in the nodal direction in the second
Brillouin zone (BZ) (a) and those at ($\pi,0$) (b) for various
doping levels \cite{Metallic}. (c) Doping dependence of the nodal
QP spectral weight, $Z_{\rm NQP}$.}
\end{figure}

Next, let us look at the doping evolution of the spectral
intensity of the QP in LSCO. Figure~\ref{EDC}(a) and (b) shows the
evolution of spectra at $\mathbf{k} \sim (\pi/2,\pi/2)$ and
($\pi,0$), respectively, with hole doping \cite{Metallic}. At
$\sim (\pi/2,\pi/2)$, a finite spectral weight exists at
$E_\mathrm{F}$ except for $x$ = 0, and increases with $x$ without
an abrupt change across the ``insulator"-superconductor transition
boundary of $x \sim 0.06$. Note that, even in the lightly-doped
non-superconducting $x$=0.03, a QP peak is depressed and crosses
the chemical potential in the nodal direction and forms a Fermi
``arc". This would explain the metallic behavior at high
temperatures of the lightly-doped materials \cite{AndoMetallic}.
To see the doping evolution of QP, the spectral weight is plotted
in Fig.~\ref{EDC}(c) as a function of hole doping $x$. Here, the
nodal QP weight $Z_{\rm NQP}$ is defined as the peak intensity of
the EDC at $k_F$ in Fig.~\ref{EDC}(a). In the underdoped regime,
$Z_{\rm NQP}$ increase rapidly with hole doping, exibiting an
apparent similarity to the doping evolution of the Drude weight in
the optical conductivity \cite{uchida1}. The observed finite
spectral weight at $x$=0.03 is consistent with the optical study
of lightly-doped LSCO, which has shown that the Drude weight is
finite already in $x$ = 0.03 \cite{dumm}.

\subsection{ Fermi Surface and Two-dimensional Band Structure }
\begin{figure*}
\includegraphics[width=14cm]{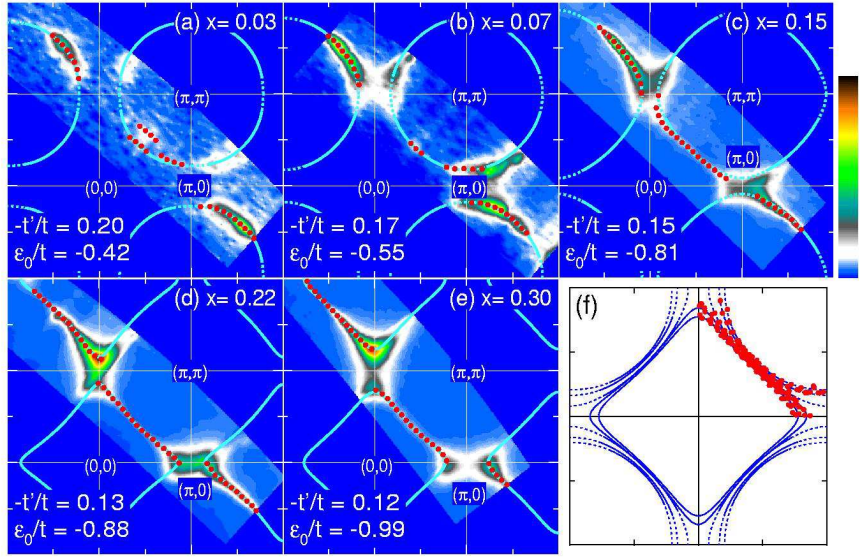}
\caption{\label{FS}(Color) Spectral weight mapping in $k$-space at
the chemical potential in LSCO. Data were taken with $h\nu$=55.5
eV. Red dots indicate $k_F$ positions determined by the peaks of
momentum distribution curves (MDC's) at the chemical potential.
The blue curves show the Fermi surface interpolated using the
tight-binding model \cite{Luttinger}. The evolution of the Fermi
surface with doping is more clearly seen in panel (f).}
\end{figure*}

%Fermi surface mapping
In order to deduce the Fermi surface or ``underlying" Fermi
surface, we have performed spectral-weight mapping at the chemical
potential in the momentum space for various doping levels as shown
in Fig. \ref{FS} \cite{Luttinger}. The spectral-weight
distribution of the overdoped $x$=0.3 as well as $x$=0.22 samples
clearly shows an electron-like Fermi surface \cite{YoshidaOD}.
Spectral weight near ($\pi$,0), a remnant ($\pi$,0) feature of the
flat band \cite{YoshidaOD}, becomes weaker in going from $x$=0.22
to $x$=0.30, indicating that, the more heavily the samples are
overdoped, the more the electronic structure becomes conventional
Fermi-liquid-like. The Fermi surface topology changes from
electron-like in $x$=0.22 to hole-like in $x$=0.15, which may
cause an electronic topological transition (ETT) \cite{Onufrieva}.
Below $x$=0.15, the nodal direction in the second Brillouin zone
(BZ) become more intense than that around ($\pi$,0), in contrast
to $x$=0.22 and 0.3, due to the superconducting or pseudogap
opening with $d$-wave symmetry in the underdoped and optimally
doped regions. In the lightly-doped $x$=0.03 [panel (a)], the
($\pi$,0) spectral weight almost disappears, and only the spectral
weight around the zone diagonal direction remains appreciable
around the chemical potential.

\begin{figure*}
\includegraphics[width=12cm]{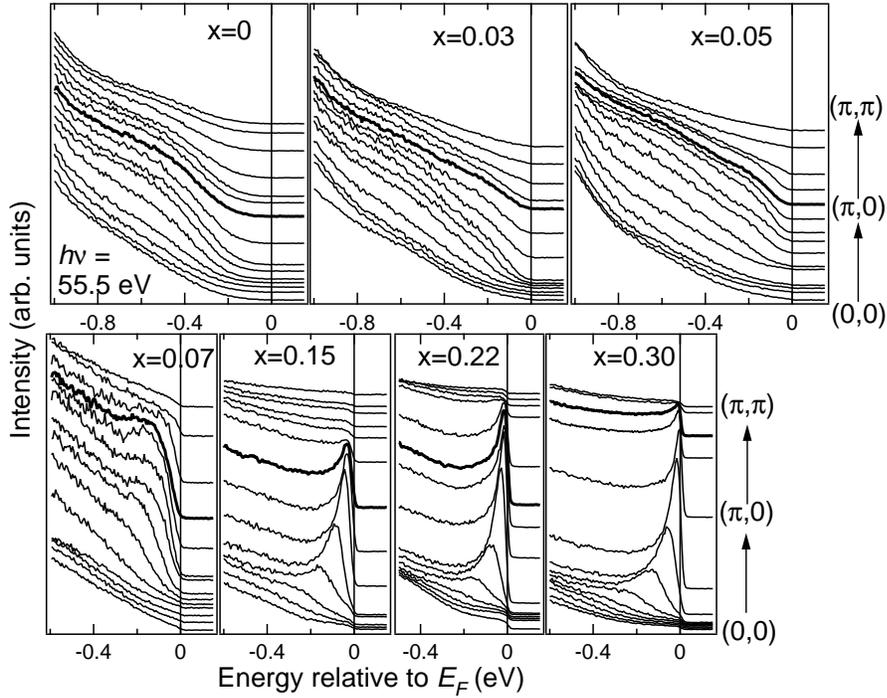}% Here is how to import EPS art
\caption{\label{EDCAll} Energy distribution curves (EDC's) along
the (0,0)-($\pi,0$)-($\pi,\pi$) lines for various doping levels in
LSCO. Thick curves represent the spectra at ($\pi,0$). The spectra
along the (0,0)-($\pi$,0) line are from the second BZ.}
\end{figure*}

\begin{figure*}
\includegraphics[width=8.5cm]{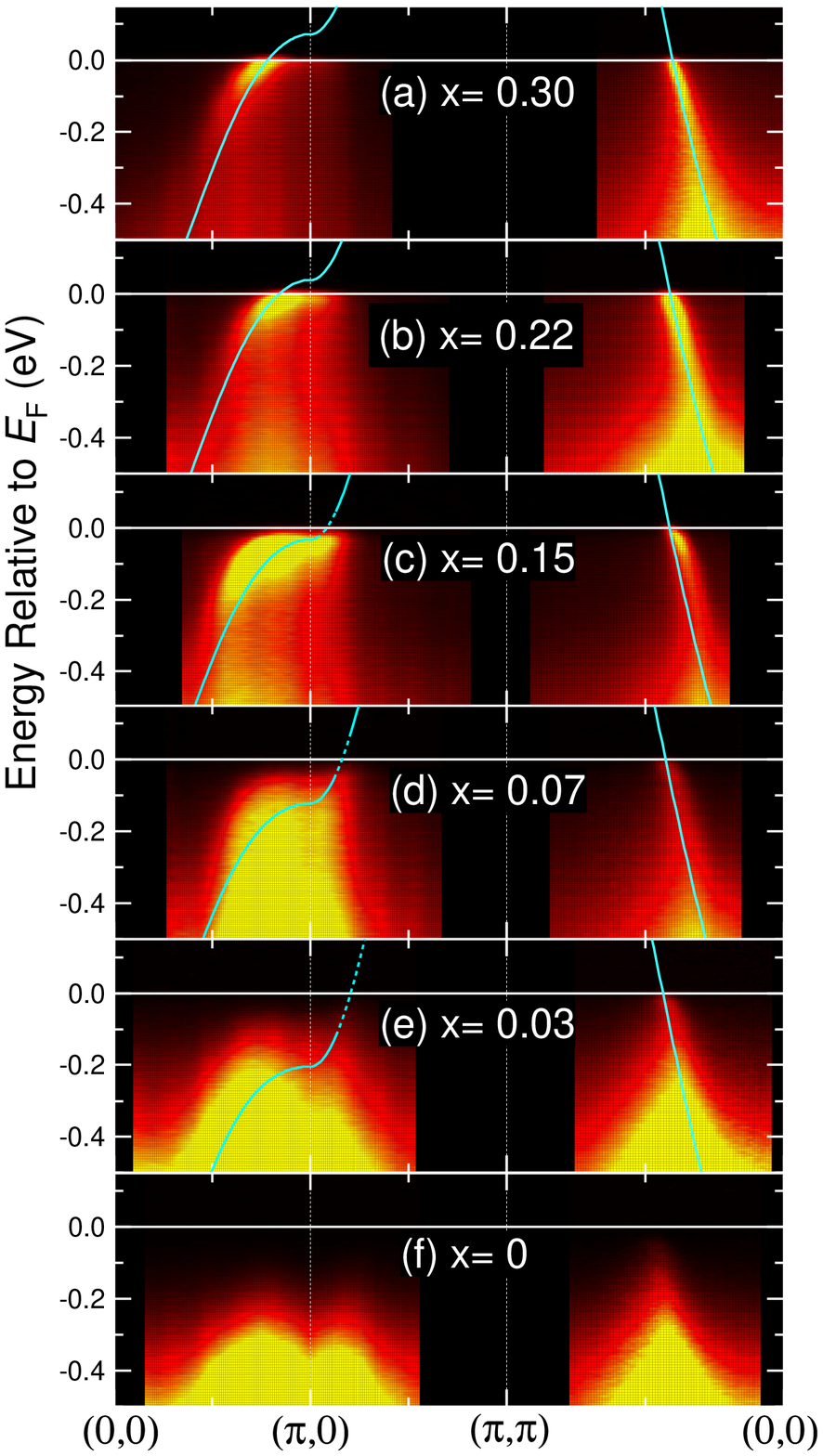}% Here is how to import EPS art
\caption{\label{disp}(Color) Intensity plot in $E$-$k$ space along
the symmetry lines (0,0)-($\pi$,0)-($\pi,\pi$)-(0,0) in LSCO. Data
were taken with $h\nu$=55.5 eV. Blue curves show the tight-binding
band dispersion with the parameters shown in Fig. \ref{FS}.}
\end{figure*}

%E-k plot
The EDC's along the $(0,0)-(\pi,0)-(\pi,\pi)$ lines for various
doping levels are shown in Fig. \ref{EDCAll} and their intensities
along the same lines are plotted in the energy-momentum $E-k$
space in Fig. \ref{disp}. One can see that the flat band feature
appears around ($\pi,0$) in the optimally and underdoped region
$x\leq$ 0.15, while the QP dispersion crosses the chemical
potential along the (0,0)-($\pi,\pi$) line. Also, the intensity
plots clearly show how the intensity at the chemical potential
around ($\pi$,0) becomes weak for $x\leq$ 0.15 as the pseudo-gap
and/or the superconducting gap opens around ($\pi$,0). One can see
from Fig. \ref{EDCAll} that the line width around ($\pi$,0)
becomes broad with decreasing $x$, indicating increasing
antiferromagnetic correlation with decreasing $x$. Also, the
``two-component" behavior is clearly seen in the EDC's of the
lightly-doped samples with $x$=0.03 and 0.05.

The ``two-component" behavior in LSCO \cite{ino2,Metallic} was
interpreted in terms of microscopic charge inhomogeneity or
stripe. According to a recent theoretical study, a mixed phase of
antiferromagnetic and superconducting states has been proposed to
explain the characteristic ``two-component" behavior in LSCO
\cite{mayr}. At the same time, a new way to interpret the
``two-component" spectral feature in terms of polaronic effects
has also emerged \cite{KMShenPRL,Nagaosa,Rosch}. In the new
picture, the high energy $\sim$ 0.5 eV have been interpreted as
polaronic side bands. These two interpretations may be related to
each other because it has been predicted that phase separation
between the insulating and metallic phases occurs under a certain
regime of electron-phonon coupling strengths \cite{capone}.

%TB fitting
The Fermi surfaces obtained from the two-dimensional tight-binding
fits are superimposed on the spectral weight mapping in Fig.
\ref{FS}. The experimental $k_{F}$'s have been fitted to the
two-dimensional (2D) single-band tight-binding (TB) model
\begin{eqnarray}
\varepsilon_k&=&-2t[\cos(k_xa)+\cos(k_ya)]-4t^\prime\cos(k_xa)\cos(k_ya)\nonumber\\
&&-2t^{\prime\prime}[\cos(2k_xa)+\cos(2k_ya)]+\varepsilon_0,
\end{eqnarray}
as shown by blue curves. Here, $t$, $t^\prime$ and
$t^{\prime\prime}$ are the first, second and third nearest
neighbor transfer integrals between Cu sites. We have assumed
constant $t=0.25$ eV and relationship
$-t^{\prime\prime}/t^\prime=1/2$ for all the doping levels, and
regarded $t^\prime$ and $\varepsilon_0$ as adjustable parameters.
The Fermi surface could be well fitted by the TB model, although
some misfit can be seen for the kink structure, which arises from
coupling to phonons as discussed in Sec. 4 \cite{LanzaraKink,
XJZhouUniversalVF} and the extremely flat band dispersion around
($\pi$,0) in the underdoped region as shown in Fig.\ref{disp}. In
Fig. \ref{FS}(f), the deduced Fermi surfaces for various $x$'s are
replotted, clearly showing the change of the Fermi surface shape
and topology with $x$. The TB dispersions with the same parameters
are superimposed on the color mapping in Fig. \ref{disp}. Overall
features of the Fermi surfaces and band dispersions are in good
accordance with the results of the local-density approximation
(LDA) band-structure calculations \cite{bansilLSCO}.

\begin{figure*}
\includegraphics[width=16cm]{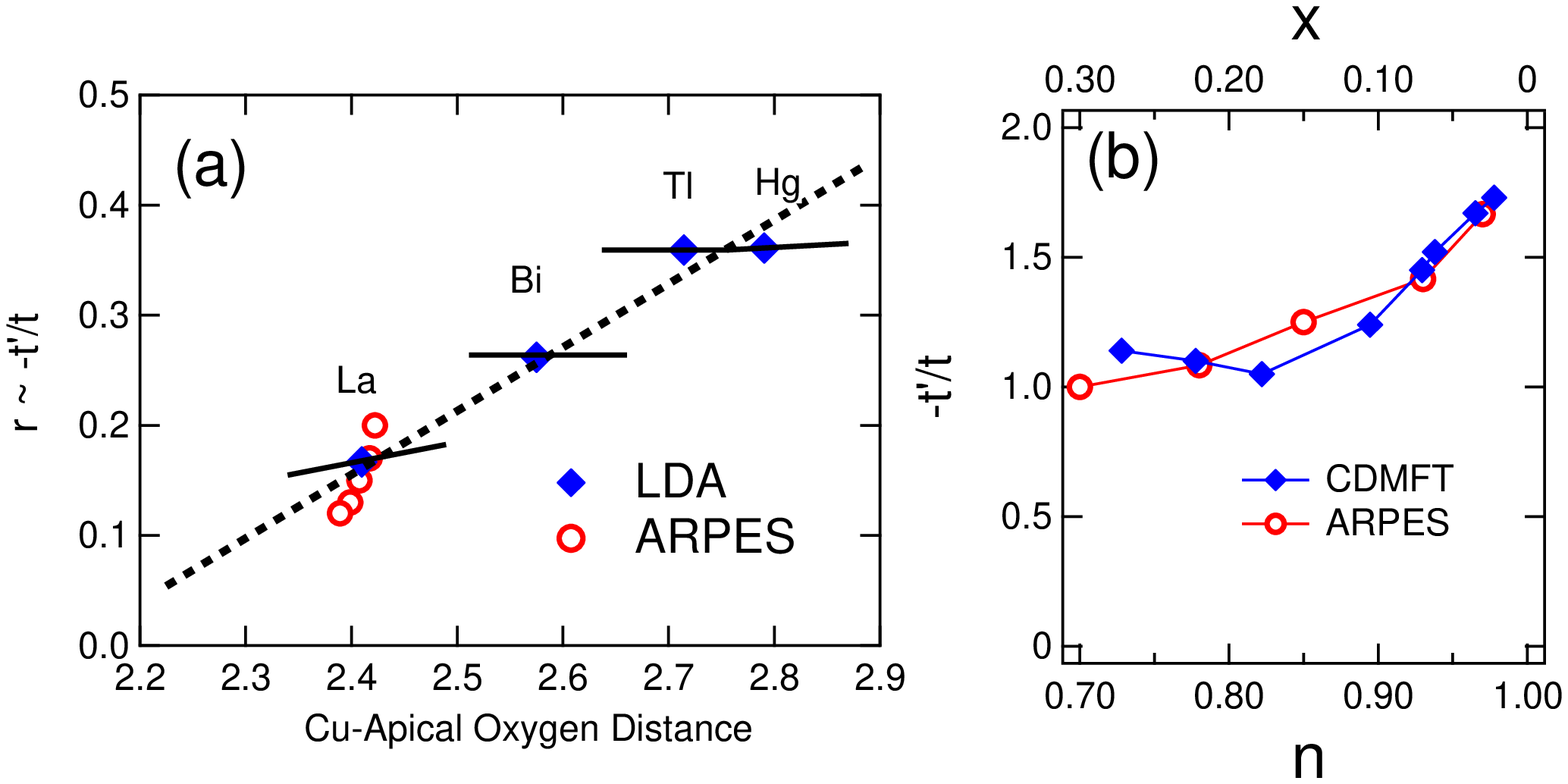}% Here is how to import EPS art
\caption{\label{tprime}(Color online) Comparison of the
$-t^{\prime}/t$ values deduced from the tight-binding fit to the
ARPES data with theoretical predictions. (a) Comparison between
$-t^\prime/t$ calculated by LDA \cite{pavarini}(filled diamond)
and those estimated from ARPES (circle), plotted as a function of
the Cu-apical oxygen distance \cite{radaelli}. The largest and
smallest $-t^\prime/t$ in the ARPES data corresponds to those of
$x$=0.03 and $x$=0.30 samples, respectively. (b) Comparison of
$t^{\prime}/t$ estimated from ARPES Fermi surface and those
calculated using CDMFT \cite{kotliar}. The theoretical values have
been normalized to the bare (i.e., input) $t^\prime/t$, while the
experimental values have been normalized to the values at $x$=0.3.
$n=1-x$ is the electron number.}
\end{figure*}

%t'
The obtained TB parameters $-t^\prime/t$ shown in Fig. \ref{FS}
exhibit a clear doping dependence. The increase of $-t^\prime/t$
with decreasing $x$ can be qualitatively explained by the increase
of the Cu-apical oxygen distance with decreasing $x$
\cite{radaelli}. In Fig.\ref{tprime}(a), the $-t^\prime/t$ values
are plotted as a function of the Cu-apical oxygen distance
\cite{radaelli} and compared with those of the LDA calculation by
Pavarini \textit{et al.} \cite{pavarini}. The obtained
$-t^\prime/t$ agree rather well with the results of the LDA
calculations $-t^\prime/t\sim$ 0.15 \cite{pavarini}. The LDA
calculation has indicated that the larger Cu-apical oxygen causes
the larger $-t^\prime/t$, which is consistent with the trend in
the obtained $-t^\prime/t$ values deduced here. The present result
is also consistent with the cellular dynamical mean field theory
(CDMFT) calculation \cite{kotliar}, in which the momentum
dependence of the self-energy has been taken into account by using
a cluster instead of a single-impurity atom in DMFT. As shown in
Fig. \ref{tprime}(b), the CDMFT calculation has indicated that
increasing correlation effects in the underdoped region is
reflected on the effective increase of the $-t^{\prime}/t$ value.
It therefore seems that both the structural effects as predicted
by the LDA calculation and the correlation effects as predicted by
the CDMFT calculation increase $-t^\prime/t$ as $x$ decreases and
explain the ARPES results.

\begin{figure*}
\includegraphics[width=14cm]{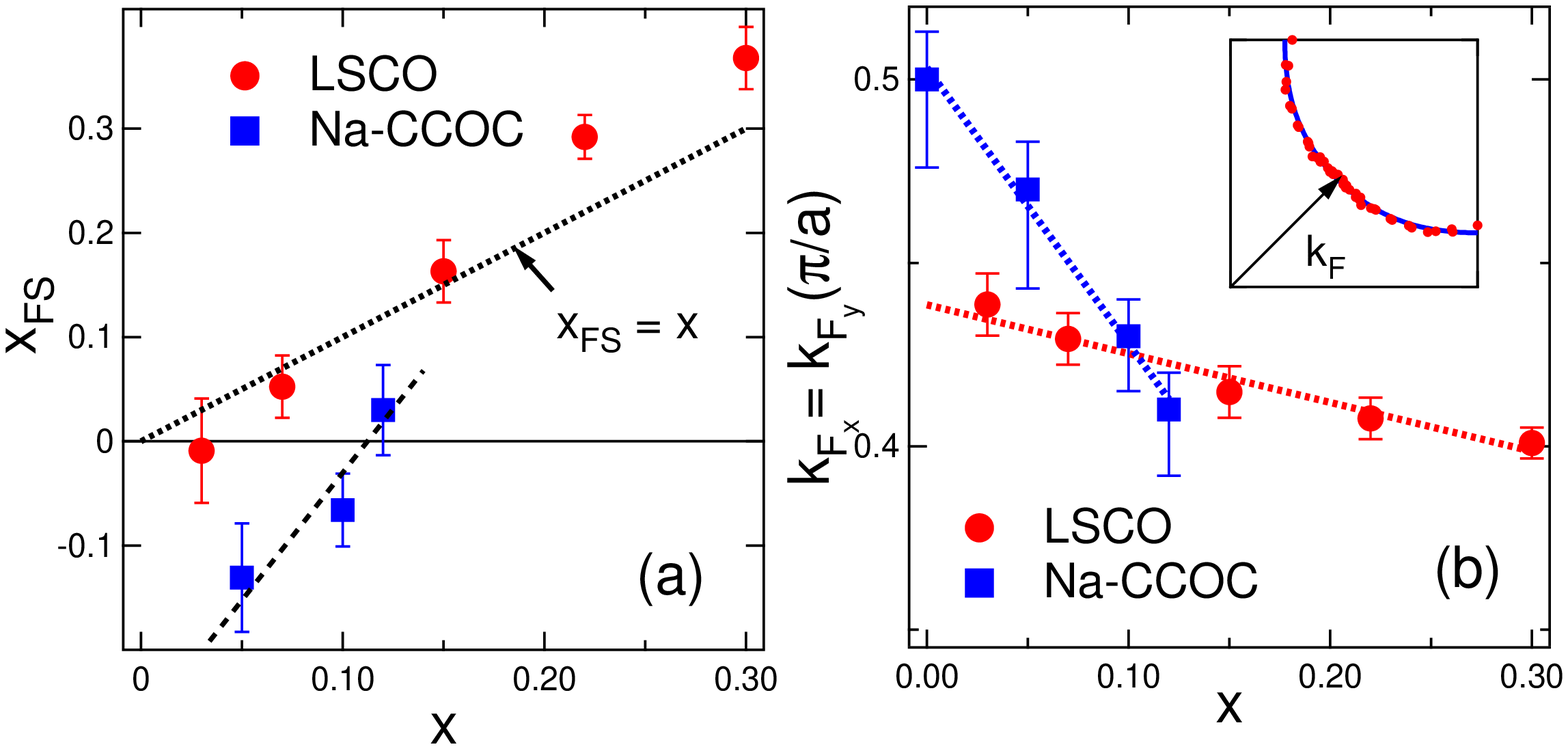}
\caption{\label{kF}(Color online) Fermi surface area and shape of
LSCO as functions of doping \cite{Luttinger}. Data for Na-CCOC
\cite{KMShenScience} are also plotted. (a) Doping dependence of
the hole number $x_{FS}$ deduced from the Fermi surface area.
Luttinger's sum rule $x_{FS}=x$ is shown for comparison. (b) Fermi
surface position in the nodal direction. }
\end{figure*}

%Luttinger's sum rule
As for Luttinger's sum rule, the underlying ``Fermi surface" of
LSCO determined from the low energy spectral weight of ARPES
approximately fulfills it remarkably well even down to the
lightly-doped region  as shown in Fig. \ref{kF}(a)
\cite{Luttinger}. This is in strong contrast with the results on
Na-CCOC, which shows a clear deviation from Luttinger's sum rule
in the underdoped region \cite{KMShenScience}. These differences
between LSCO and Na-CCOC may be related to the difference in the
$t^{\prime}/t$ value and/or to the difference in the
electron-lattice coupling strength. In relation to the area of the
Fermi surface, the shape of the Fermi surface may be characterized
by the position of the Fermi surface along the nodal direction. As
shown in Fig. \ref{kF} (b), the same position in Na-CCOC shows a
strong doping dependence and is extrapolated to ($\pi$/2,$\pi$/2)
in the $x$=0 limit, whereas in LSCO, the limiting value is off the
($\pi$/2,$\pi$/2) point \cite{Luttinger}. This may be related to
the observation that the chemical potential approaches the top of
the lower Hubbard band in Na-CCOC as the doping decreases whereas
the chemical potential stays away from the lower Hubbard band in
LSCO \cite{inoCP,KMShenPRL,Yagi}.

\subsection{Large Pseudogap and Small Pseudogap}
\begin{figure*}
\includegraphics[width=12cm]{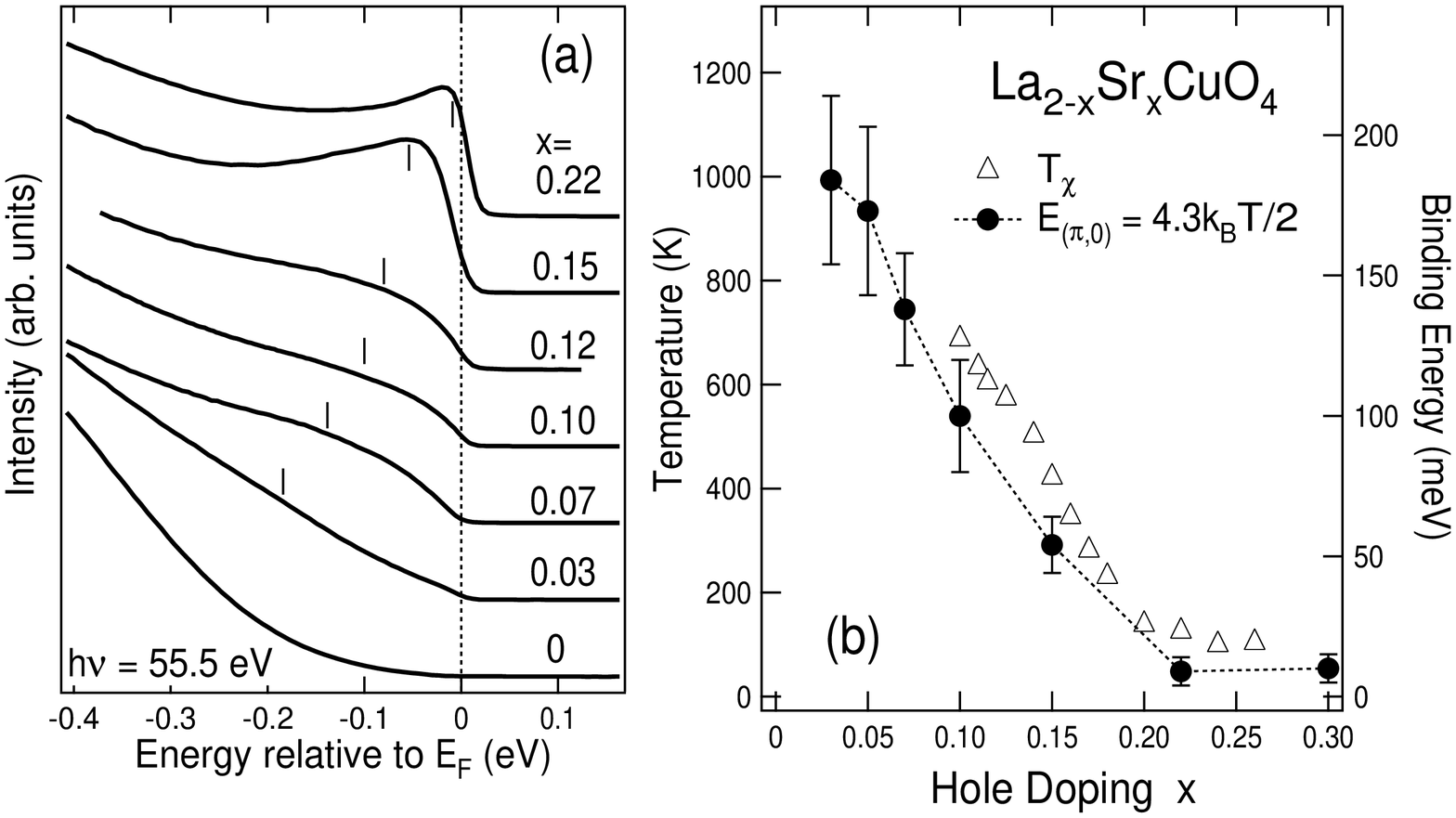}% Here is how to import EPS art
\caption{\label{Lgap} Angle-integrated photoemission spectra and
the large pseudogap. (a) Spectrum for each doping. Vertical bars
represent the binding energy of the flat band. (b) Doping
dependence of the binding energy of the flat band is compared with
the characteristic temperature $T_\chi$ where the magnetic
susceptibility $\chi(T)$ takes a maximum \cite{nakano1}.}
\end{figure*}

There are two kinds of ``pseudogaps" in the high-$T_c$ cuprates
called a ``small pseudogap" and a ``large pseudogap". First, we
discuss the relationship between the high-energy pseudogap
behaviors, which have been observed as a characteristic
temperature in the magnetic susceptibility $\chi(T)$
\cite{nakano1} and the Hall coefficient $R_H(T)$ \cite{hwang}, and
a characteristic energy in the photoemission spectra. As shown in
Fig. \ref{Lgap}(b), the doping dependence of the flat band
position $E_{(\pi,0)}$ agrees with that of $T_\chi$ below which
$\chi(T)$ decreases from the maximum value, if the relationship
for the $d$-wave mean-field relation $E_{(\pi,0)}=4.3k_BT/2$ is
assumed. Figure \ref{Lgap}(a) shows angle-integrated spectra
obtained by integrating ARPES spectra in the second BZ, where the
$E_{(\pi,0)}$ is indicated by vertical bars. Since the flat band
gives a singularity in the density of states (DOS), the
corresponding maximum is seen at $E_{(\pi,0)}$ in the
angle-integrated spectra \cite{inoAIPES,satoLSCO}. This energy
scale may be related to the strong suppression of the spectral
weight over an extended energy range, particularly prominent near
($\pi$,0). These results are consistent with the theoretical
prediction of ETT that $E_{(\pi,0)}$ is a crucial factor to
determine the quantum critical point \cite{Onufrieva}.

\begin{figure*}
\includegraphics[width=11cm]{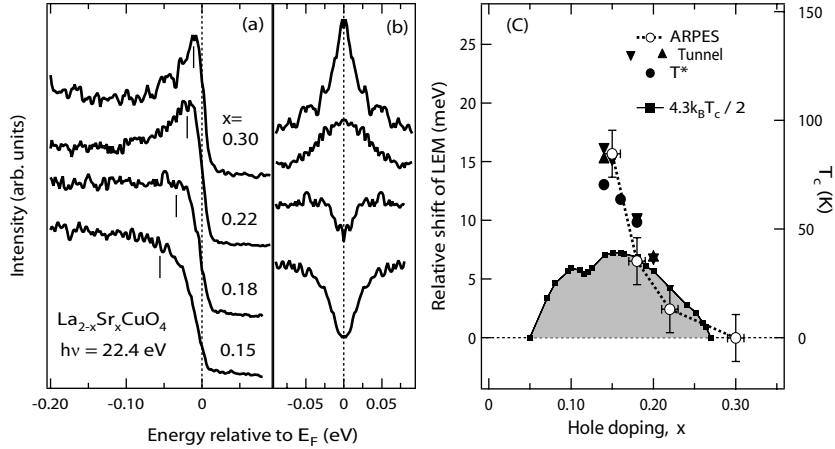}% Here is how to import EPS art
\caption{\label{pi0gap} ARPES spectra on the Fermi surface in the
antinodal region. (a) EDC for each doping. (b) Symmetrized spectra
of (a). (c) Relative shift of the leading-edge midpoint (LEM) in
the ARPES spectra in (a). The LEM shift is compared with the
$T_{\mathrm c}$ (filled square)  and the gap measured by scanning
tunnel spectroscopy (filled triangles) \cite{nakano2}, assuming
the relationship $E=4.3k_B T$ between energy and temperature
\cite{won}.}
\end{figure*}

Next, we discuss the relationship between the small pseudo gap
behaviors of the thermodynamic and transport properties and the
low-energy spectral feature of ARPES spectra. Figure
\ref{pi0gap}(a) shows ARPES spectra on the Fermi surface in the
antinodal direction. In order to clearly show the gap opening, the
spectra in Fig. \ref{pi0gap}(a) have been symmetrized with respect
to $E_{\mathrm F}$ as shown in Fig. \ref{pi0gap}(b). For $x$= 0.3
and 0.22, the symmetrized spectra show a peak at the chemical
potential in contrast to those for $x$=0.15, corresponding to the
normal Fermi-liquid behavior of the transport results
\cite{takagi}. Figure \ref{pi0gap}(c) shows the doping dependence
of the relative energy of the leading edge midpoint (LEM).
Although the LEM in $x$=0.22 is above the Fermi level in spite of
$T_c >10$K, it is by $\sim(2-3)$ meV below that of the
non-superconducting sample $x$=0.3, implying the existence of a
small gap. Note that, when the QP is at the Fermi level, the LEM
would be above the Fermi level due to the energy broadening under
the finite energy resolution. As seen in Fig. \ref{pi0gap}(c), the
observed gap well corresponds to the small pseudogap observed by
the scanning tunneling measurements.

\section{Effects of Electron-Phonon Coupling}

It has been recognized from the very beginning of the high-$T_c$
research that many-body effects are key to understanding cuprate
physics. Due to its proximity to the antiferromagnetic Mott
insulating state, electron-electron interactions are extensively
discussed in the
literature\cite{DamascelliReview,CampuzanoReview}. In this
Section, we shall briefly describe the present status of our
understanding of electrons interacting with phonons. A more
comprehensive review is presented elsewhere\cite{XJZhouReview}.

\subsection{Electron-Phonon coupling along the Nodal Direction}

\begin{figure}[tbp]
\begin{center}
\includegraphics[width=1.0\linewidth,angle=0]{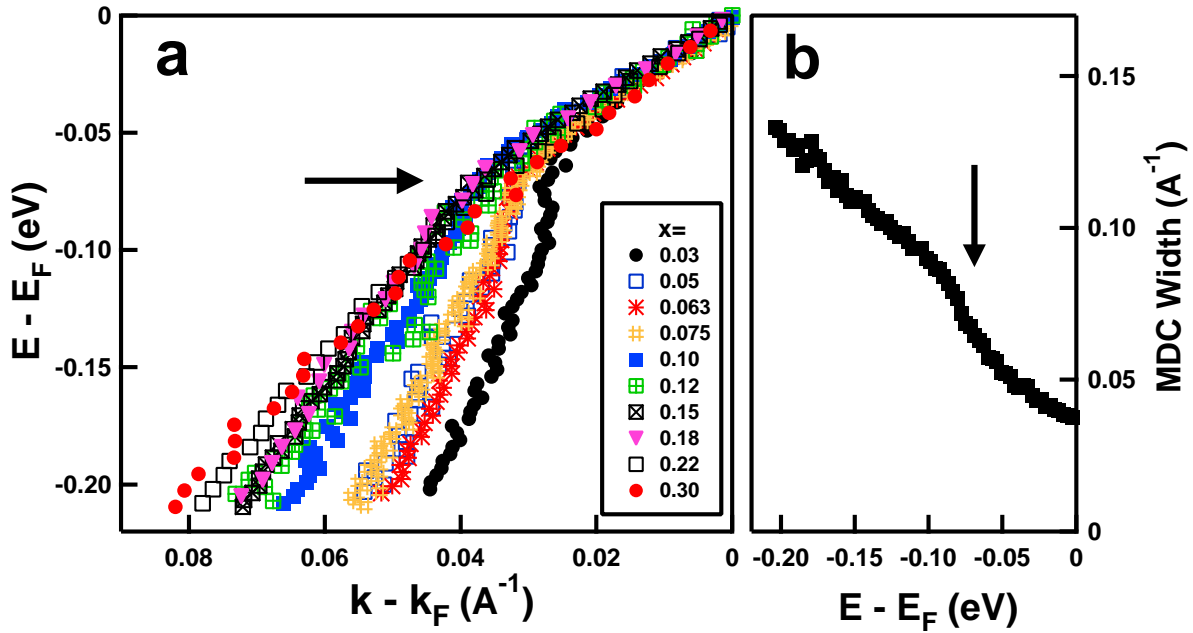}
\end{center}
\caption{(Color online) Doping dependence of the nodal QP in
LSCO\cite{XJZhouUniversalVF}. (a) Dispersion of LSCO with various
doping levels (x=0.03 to 0.30) measured at 20 K along the
(0,0)-($\pi$,$\pi$) nodal direction.  (b). Scattering rate as
measured by MDC width (full-width-at-half-maximum, FWHM) of the
LSCO (x = 0.063) sample measured at 20 K.}\label{UVelocity}
\end{figure}

The {\it d}-wave superconducting gap is zero along the nodal
direction. As shown in Fig.\ref{UVelocity}(a), the energy-momentum
dispersion curves from MDC method exhibit an abrupt slope change
(``kink") near 70 meV. The kink is accompanied by an accelerated
drop in the MDC width at a similar energy scale (Fig.
\ref{UVelocity}(b)). The existence of the kink has been well
established as ubiquitous in hole-doped cuprate
materials\cite{BogdanovKink,LanzaraKink,KaminskiKink,PJohnsonKink,BorisenkoNodalKink,XJZhouUniversalVF,GweonIsotope}:

1.  It is present in various hole-doped cuprate materials,
including Bi2212, Bi$_2$Sr$_2$CuO$_6$ (Bi2201), LSCO and others.
The energy scale (in the range of 50-70 meV) at which the kink
occurs is similar for various systems.

2.  It is present both below T$_c$ and above T$_c$.

3.  It is present over an entire doping range (Fig.
\ref{UVelocity}(a)). The kink effect is stronger in the underdoped
region and gets weaker with increasing doping.

While there is a consensus on the data, the exact meaning of the
data is still under discussion. The first issue concerns whether
the kink in the normal state is related to an energy scale. Valla
\textit{et al.} \cite{VallaScience} argued that the system is
quantum critical and thus has no energy scale, even though a band
renormalization is present in the data. Since their data do not
show a sudden change in the scattering rate at the corresponding
energy, they attributed the kink in Bi2212 above T$_c$ to the
marginal Fermi liquid (MFL) behavior without an energy
scale\cite{PJohnsonKink}. Others believe the existence of energy
scale in the normal and superconducting states has a common
origin, i.e., coupling of quasiparticles with low-energy
collective excitations
(bosons)\cite{BogdanovKink,LanzaraKink,KaminskiKink}. The sharp
kink structure in dispersion and concomitant existence of a drop
in the scattering rate which is becoming increasingly clear with
the improvement of signal to noise in the data, as exemplified in
underdoped LSCO (x=0.063) in the normal state (Fig.
\ref{UVelocity}(b))\cite{XJZhouUniversalVF}, are apparently hard
to reconcile with the MFL behavior.

A further issue concerns the origin of the bosons involved in the
coupling, with a magnetic resonance mode
\cite{KaminskiKink,PJohnsonKink} and optical phonons
\cite{LanzaraKink} being possible candidates considered. The
phonon interpretation is based on the fact that the sudden band
renormalization (or ``kink") effect is seen for different cuprate
materials, at different temperatures, and at a similar energy
scale over the entire doping range \cite{LanzaraKink}. For the
nodal kink, the phonon considered in the early work was the
half-breathing mode, which shows an anomaly in neutron experiments
\cite{PintschoviusBreathingMode,McQueeneyBreathingMode}. Unlike
the phonons, which are similar in all cuprates, the magnetic
resonance (at correct energy) is observed only in certain
materials and only below T$_c$. The absence of the magnetic mode
in LSCO and the appearance of the magnetic mode only below T$_c$
in some cuprate materials are not consistent with its being the
cause of the universal presence of the kink effect. Whether the
magnetic resonance can cause any additional effect is still an
active research topic \cite{KeeMagneticMode,AbanovMagneticMode}.

While the nodal data clearly reveal the presence of coupling to
collective modes with well-defined energy scale, there are a
couple of peculiar behaviors associated with the doping evolution
of the nodal dispersion.  As seen from Fig. \ref{UVelocity}(a),
the low-energy dispersion and velocity from the MDC method is
insensitive to doping over the entire doping range, while the high
energy velocity increases with decreasing
doping.\cite{XJZhouUniversalVF}. In terms of conventional
electron-phonon coupling, if one considers that the ``bare band"
does not change with doping but the electron-phonon coupling
strength increases with decreasing doping, as it is probably the
case for LSCO,  one would expect that the low energy dispersion
and velocity show strong doping dependence, while the high-energy
ones converge. This is opposite to the experimental observations.
Moreover, considering the electron-electron interaction gets
stronger with decreasing doping,  according to conventional
wisdom, this would result in a larger effective mass and smaller
velocity, which is again opposite to experimental observation
(Fig. \ref{UVelocity}(a)). These anomalies indicate a potential
deviation from the standard Migdal-Eliashberg theory and the
possibility of a complex interplay between electron-electron and
electron-phonon interactions. This phenomenon is a hint of
polaronic effect where the traditional analysis fails.  Such a
polaron effect gets stronger in deeply underdoped system even
along the nodal direction. It is noted that, while the low energy
velocity is rather ``universal" via MDC analysis, the low energy
dispersions derived from EDC and MDC methods are rather different
\cite{Nagaosa}. The difference is most pronounced at low doping
where the coupling is strong. In this case, the EDC-derived
velocity exhibits considerable doping dependence. While for weakly
interacting systems MDC and EDC are expected to yield similar
dispersion relations from linearly dispersing bands, strong
coupling features which cause substantial renomaliations may cloud
EDC and MDC analysis. The origin of the difference between MDC and
EDC-derived dispersions remains a matter to be further explored.

\subsection{Multiple Modes in the Electron Self-Energy}

In conventional superconductors, the successful extraction of the
phonon spectral function, or the Eliashberg function,
$\alpha^{2}F(\omega)$,  from electron tunneling data played a
decisive role in cementing the consensus on the phonon-mediated
mechanism of superconductivity \cite{Rowell}. For high temperature
superconductors, the extraction of the bosonic spectral function
can provide fingerprints for more definitive identification of the
nature of bosons involved in the coupling.

In principle, the ability to directly measure the dispersion, and
therefore, the electron self-energy, would make ARPES the most
direct way of extracting the bosonic spectral function. This is
because, in metals, the real part of the electron self-energy
Re$\Sigma$ is related to the Eliashberg function
$\alpha^{2}F(\Omega;\epsilon,\mathbf{\hat{k}})$ by:
\begin{equation}
\mathrm{Re}\Sigma(\mathbf{\hat{k}},\epsilon;T)=\int_{0}^{\infty}d\Omega\alpha^{2}F(\Omega;\epsilon,\mathbf{\hat{k}})K\left(\frac{\epsilon}{kT},\frac{\hbar\Omega}{kT}\right)\,,\label{eq:ReSigma}
\end{equation}

\noindent where

\begin{equation}
K(y,z)=\int_{-\infty}^{\infty}dx\frac{2z}{x^{2}-z^{2}}f(x+y)\,,\label{eq:MEMk}
\end{equation}
with $f(x)$ being the Fermi distribution function. Such a relation
can be extended to any electron-boson coupling system and the
function  $\alpha^{2}F(\omega)$ then describes the underlying
bosonic spectral function. Unfortunately, given that the
experimental data inevitably have noise, the traditional
least-square method to invert an integral problem is
mathematically unstable.

Very recently,  Shi \textit{et al.} have made an important advance
in extracting the Eliashberg function from ARPES data by employing
the maximum entropy method (MEM) and successfully applied the
method to Be surface states \cite{JRShiMEM}.  The MEM approach
\cite{JRShiMEM} is advantageous over the least squares method in
that: (i) It treats the bosonic spectral function to be extracted
as a probability function and tries to obtain the most probable
one. (ii) More importantly, it is a natural way to incorporate the
{\it priori} knowledge as a constraint into the fitting process.
In practice, to achieve an unbiased interpretation of data,  only
a few basic physical constraints to the bosonic spectral function
are imposed: (A) It is positive. (B) It vanishes at the limit
$\omega$$\rightarrow$0. (C) It vanishes above the maximum energy
of the self-energy features. As shown in the case of Be surface
state, this method is robust in extracting the Eliasberg function
\cite{JRShiMEM}.

\begin{figure}[htb]
\includegraphics[width=.5\textwidth]{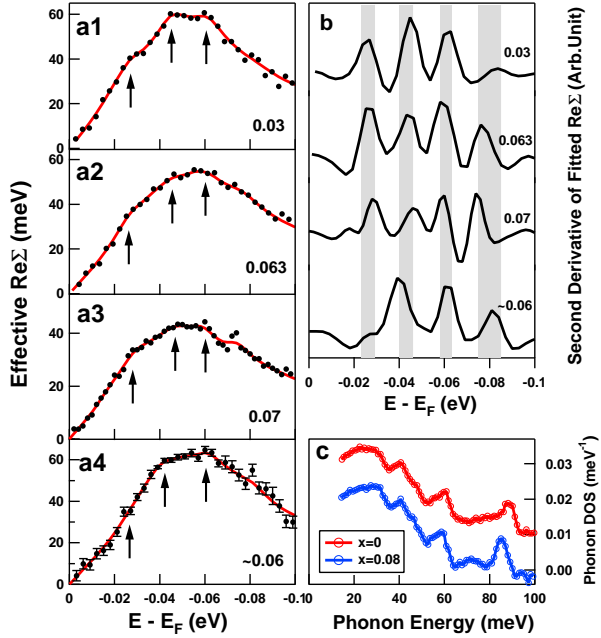}
\caption{(Color online) Multiple modes coupling in the electron
self-energy in LSCO\cite{XJZhouMultipleMode}.   (a) Effective real
part of the electron self-energy for LSCO x=0.03 (a1), 0.063 (a2),
0.07 (a3)and $\sim$0.06 (a4) samples.  (b). Second-order
derivative of the calculated Re$\Sigma$. The four shaded areas
correspond to energies of 23-29, 40-46, 58-63 and 75-85 meV where
the fine features fall in. (c) Phonon density of states
$F(\omega)$ for LSCO $x=0$ (red) and $x=0.08$ (blue) measured by
neutron scattering \cite{McQueeneyDOS}. }\label{Multimode}
\end{figure}

Initial efforts have been made to extend this approach to
underdoped LSCO and evidence for electron coupling to several
phonon modes has been revealed \cite{XJZhouMultipleMode}. As seen
from Fig. \ref{Multimode}, from both the electron self-energy(Fig.
\ref{Multimode}(a)), and the derivative of their fitted curves
((Fig. \ref{Multimode}(a)), one can identify two dominant features
near $\sim$ 40 meV and $\sim$60 meV. In addition, two addition
modes may also be present near $\sim$25meV and $\sim$75 meV
\cite{XJZhouMultipleMode}.  The multiple features in Fig.
\ref{Multimode}(b) show marked difference from the magnetic
excitation spectra measured in LSCO which is mostly featureless
and doping dependent \cite{HaydenTranquada}. In comparison, they
show more resemblance to the phonon density-of-states (DOS),
measured from neutron scattering on LSCO (Fig.\ref{Multimode}(c))
\cite{McQueeneyDOS}, in the sense of the number of modes and their
positions. This similarity between the extracted fine structure
and the measured phonon features favors phonons as the bosons
coupling to the electrons. In this case, in addition to the
half-breathing mode at 70$\sim$80 meV that we previously
considered strongly coupled to electrons \cite{LanzaraKink}, the
present results suggest that several lower energy optical phonons
of oxygens are also actively involved.

We note that, in order to be able to identify fine structure in
the electron self-energy, it is imperative to have both high
energy resolution and high statistics \cite{ZhouValla}.  These
requirements have made the experiment highly challenging because
of the necessity to compromise between two conflicting
requirements for the synchrotron light source: high energy
resolution and high photon flux. Further improvements in
photoemission experiments will likely enable a detailed
understanding of the boson modes coupled to electrons, and provide
critical information for the pairing mechanism.

\section{Relation to thermodynamic and transport properties}
The thermodynamics and transport properties of the
high-$T_{\textrm{c}}$ cuprates are fundamentally affected by the
strong electron correlation near the filling-control Mott
transition, particularly by the opening of the pseudogap as
implied by Fig.\ref{PG}. The electronic specific heat coefficient
$\gamma$ decreases with decreasing hole concentration in the
underdoped region \cite{momono}. The electrical resistivity shows
unconventional electrical resistivity such as the $T$-liner
temperature dependence in the optimally doped region \cite{takagi}
and the metallic behavior with resistivity well exceeding the
Ioffe-Regel limit in the lightly doped region \cite{AndoMetallic}.
In order to elucidate the origin of these unconventional
phenomena, in this Section, we shall attempt to explain the doping
dependence of the electronic specific heat coefficient $\gamma$,
superfluid density, the electrical resistivity and the Hall
coefficient on the basis of the ARPES results.

\begin{figure*}
\includegraphics[width=14cm]{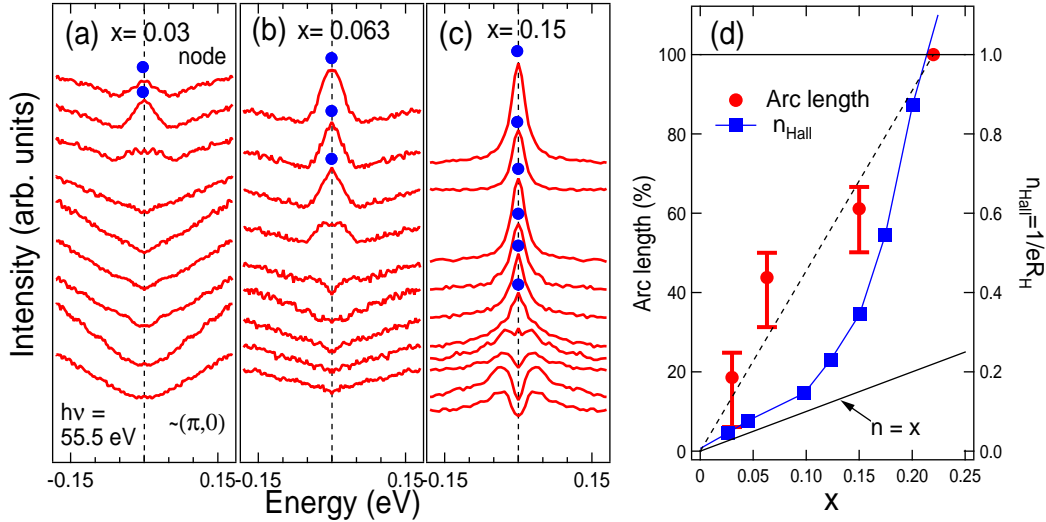}
\caption{\label{arc}(Color online) Symmetrized EDC's along the
(underlying) Fermi surface for $x$=0.03(a), $x$=0.063(b) and
$x$=0.15(c). Panel (d) shows the arc length ratio defined by the
fraction of the Fermi surface length which shows a QP peak as in
the symmetrized EDC's. Curves for $n = x$ and $n_{\rm Hall} =
1/eR_H$ \cite{takagi} are also shown.}
\end{figure*}

\subsection{Electronic specific heat}
The electronic specific heat coefficient $\gamma$ of a Fermi
liquid is given by $\gamma=(\pi^2k_B^2/3) N(0)^*$, where $N(0)^*$
is the QP density of states (DOS) at the chemical potential. In a
single-band system like single-layer cuprates, each momentum
$\mathbf{k}$ contributes one QP, if the system is a Fermi liquid.
Therefore, by using the Fermi velocity $v_F$ of the QP band along
the Fermi surface, one can deduce the two-dimensional QP DOS
$N(0)^*$ using the formula $N(0)^*=(1/2\pi^2)\int ds / \hbar v_F$,
where the integrations is made over the Fermi surface.

\begin{figure*}
\includegraphics[width=14cm]{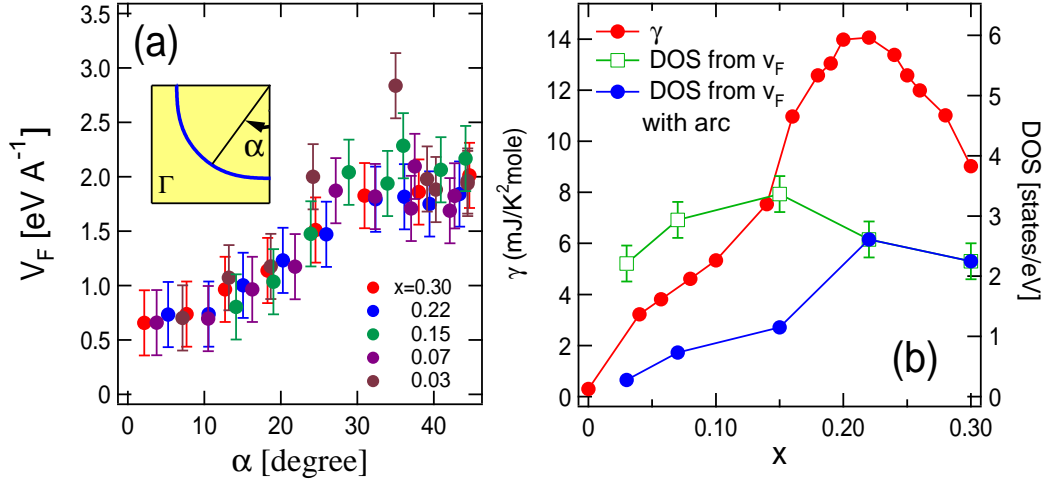}
\caption{\label{gamma}(Color online) Density of states at the
chemical potential in LSCO. (a) Fermi velocity $v_F$ along the
Fermi surface. (b) Comparison the electronic specific heat
coefficient $\gamma$ \cite{momono} and the QP DOS at the chemical
potential calculated using $v_F$ from ARPES data. The QP DOS from
ARPES with the arc length taken into account is also shown.}
\end{figure*}

In order to take into account the effect of the pseudogap in the
present analysis, we first attempted to derive the length of the
Fermi arc. From symmetrized EDC's on the (underlying) Fermi
surface as shown in Fig. \ref{arc} (a)-(c), we have defined the
arc length by the portion of the (underlying) Fermi surface
showing a single QP peak at the chemical potential. As shown in
Fig. \ref{arc}(d), one can see that the arc length evolves with
hole doping systematically and eventually it becomes a fully
extended (100\%) closed Fermi surface in overdoped $x$=0.22. The
carrier number defined by $n_{\rm Hall}=1/eR_H$ are compared with
the arc length in Fig. \ref{arc}(d). The difference between
$n_{\rm Hall}$ and the arc length may come from the fact that
$n_{Hall}$ is affected by the curvature of the Fermi surface, too
\cite{Ong}. Note, however, that due to the finite energy
resolution, closely spaced two QP peaks may not be resolved in the
symmetrized EDC's and that the arc length may be overestimated.

Figure \ref{gamma}(a) shows the Fermi velocity $v_F$ for various
doping levels determined from the observed QP dispersions as a
function of Fermi-surface angle $\alpha$. Here, we have used MDC
for determining $v_F$ in a similar manner as in Fig.
\ref{UVelocity}, while an alternative way is to use EDC's
\cite{MannellaCMR}. One can see that the $v_F$ remains relatively
unchanged with hole doping not only in the node direction
($\alpha=45^\circ$; see Fig. \ref{UVelocity}) but also on the
entire Fermi surface. In Fig. \ref{gamma}(b), the electronic
specific heat coefficient $\gamma$ calculated using the $v_F$
values thus deduced is compared with the $\gamma$ of LSCO samples
where the $T_{\textrm{c}}$ was suppressed by Zn-doping
\cite{momono}. The $N(0)^*$ calculated from ARPES has a maximum
around $x$=0.15 because the flat band near ($\pi,0$) becomes
closest to the Fermi level. If the system is a Fermi liquid,
$N(0)^*$ from ARPES and $\gamma$ should agree with each other.
However, $\gamma$ decreases much faster than $N(0)^*$ from ARPES
with decreasing $x$, indicating that the QP density at the
chemical potential is indeed depleted in the underdoped materials
and demonstrating that the opening of the pseudogap around
($\pi,0$) affects the low-energy QP excitations. If the shrinking
arc length is taken into account, one can qualitatively reproduce
the decrease of $\gamma$ with decreasing $x$. As for the absolute
values, however, the experimental values are larger than those
deduced from ARPES by a factor of $\sim$2. The origin of this
discrepancy is not known at present.

\subsection{Superfluid density}
\begin{figure*}
\includegraphics[width=8cm]{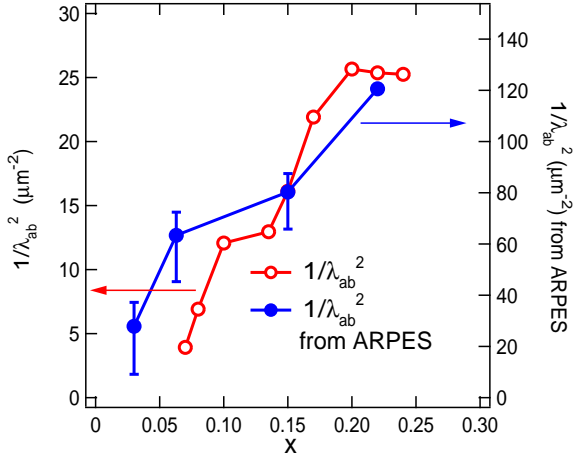}
\caption{\label{superfluid}(Color online) In-plane superfluid
density deduced from the magnetic penetration depth $\lambda_{ab}$
\cite{panagopoulos} compared with that estimated from the ARPES
data.}
\end{figure*}

Comparison between the ARPES spectra and the superfluid density
would also give important information about the pseudogap opening
and its effects on the superconductivity \cite{fengScience}. The
in-plane London penetration depth $\lambda_{ab}$ and the
superfluid density $\rho_s=n_s / m^*$ are related through
$1/\lambda_{ab}^2 =\mu_0 e^2 n_s/m^*$, where $\mu_0$ is the vacuum
permeability, and $n_s$ is the density of superconducting
electrons. For the density of superconducting electrons, we
assumed that $n_s$ per Cu atom is given by $(1-x)\times$(arc
length), where the doping dependence of the arc length is given in
Fig.\ref{arc}(d). Since the nodal direction rather than the
antinodal direction is expected to be the main contribution to the
in-plane superfluid density due to the lighter effective mass,
$m^*$ was approximated by $m^*=\hbar k_F/v_F$, where $v_F$ is the
Fermi velocity in the nodal direction \cite{XJZhouUniversalVF}
(Sec. 4.1). In Fig. \ref{superfluid}, the doping dependence of
$1/\lambda_{ab}^2 $ \cite{panagopoulos} is compared with that
deduced from the ARPES data. The two curves show qualitatively
good agreement between them. Considering that $m^*$ is nearly
independent of doping \cite{XJZhouUniversalVF}, the doping
dependence of the arc length is the principal origin of the doping
dependence of $n_s$. Quantitatively, however, the
$1/\lambda_{ab}^2 $ values show a difference of factor $\sim$5,
implying that the effective density of superconducting electrons
is reduced compared with the nominal carrier number even in the
overdoped region. Also, note that the superfluid density becomes
finite only for $x>0.05$, while that estimated from ARPES shows
finite value even in $x$=0.03. Another viewpoint to explain the
reduced superfluid density is to attribute it to the reduced QP
spectral weight $Z$ rather than the enhanced $m^*$
\cite{fengScience,Paramekanti}, which also explain the doping
dependence of $1/\lambda_{ab}^2$ [see Fig.\ref{EDC}(c)]. The
overall quantitative difference may come from the effect of $Z$.

\subsection{Electrical resistivity}
\begin{figure*}
\includegraphics[width=14cm]{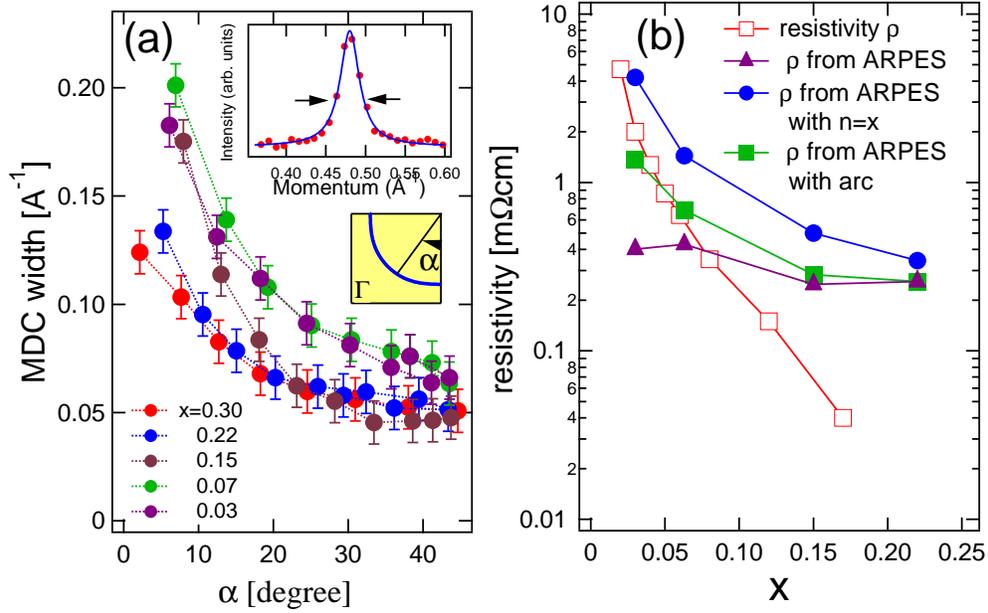}
\caption{\label{resistivity}(Color online) Electrical resistivity
of LSCO estimated from ARPES. (a) MDC width for various doping
levels plotted as a function of Fermi-surface angle $\alpha$. (b)
Comparison of the resistivity estimated using the MDC width and
the electrical resistivity at 20 K \cite{AndoMetallic}. For the
transport data for those compositions which show localization
behavior at low temperatures, the minimum resistivity at a higher
temperature is adopted.}
\end{figure*}

Because of the pseudo-gap opening around ($\pi$,0), the electronic
states around the node mainly contribute to the in-plane transport
properties in the underdoped region. Here, we have attempted to
estimate the in-plane electrical resistivity from the observed MDC
width $\Delta k=1/l_k$ [Fig. \ref{resistivity}(a)], where $l_k$ is
the mean-free-path, and the length of the ``Fermi arc" obtained in
Sec.5.1 as follows. In the simple Drude formula, the resistivity
is given by $\rho=m^*/ne^2\tau=\hbar k_F \Delta k /ne^2$. It is
often assumed that $n=x$ based on the Hall coefficient results
\cite{takagi}. In the Boltzmann equation in two-dimension, the
diagonal component of the conductivity is given by
$\sigma_{xx}=(e^2/h\pi)\int l_k \cos^2 \theta_k ds$, where the
integration is made over the Fermi surface, and $\theta_k$ is the
angle between $\mathbf{l}_k=\mathbf{v}_F \tau_k$ and the electric
field taken parallel to the $x$-axis \cite{Ong}. Using ARPES data
of high-$T_c$ cuprates, the electrical resistivity at high
temperatures, where the pseudogap disappears, has been calculated
using this method \cite{kondo}. In addition to this treatment, we
have taken into account the effect of ``Fermi arc" by reducing the
integration length on the Fermi surface.

The MDC width at the chemical potential is plotted in Fig.
\ref{resistivity}(a) as a function of Fermi-surface angle
$\alpha$. One can see that the width around the node
($\alpha$=45$^\circ$) depends on doping only weakly, which is
consistent with the weak doping dependence of the charge mobility
$\mu=e\tau/m^*=e/\hbar k_F \Delta k $ \cite{AndoMetallic}. Using
the data in Fig. \ref{resistivity}(a), the doping dependence of
the in-plane electrical resistivity $\rho$ was estimated by
applying the Drude formula with $n=x$ and the nodal MDC width, and
the Boltzmann formula with and without the Fermi arc. Then, they
are compared with the transport results \cite{AndoMetallic} in
Fig. \ref{resistivity}(b).

In all the $\rho$ estimated from ARPES, the discrepancy from the
transport data is pronounced in the over doped region. The
discrepancy stems from the wide MDC widths compared to that
expected from the transport results in the overdoped region, which
may be due to the effects of vertex correction, the possible
dominance of forward scattering \cite{Zhu} in the overdoped region
and/or the finite instrumental resolution neglected in the present
analysis. On the other hand, in the underdoped region, the $\rho$
estimated from ARPES including the effects of ``Fermi arc" or
assuming $n=x$ is in good agreement with the transport results,
while that evaluated using the entire Fermi surface is by more
than one order of magnitude smaller than the transport data. This
suggests that the pseudogap is the origin of the small carrier
number and hence the high electrical resistivity of the underdoped
samples. As a whole, the $\rho$ values estimated from ARPES with
the Fermi arc well explains the doping dependence of the transport
data of the underdoped LSCO.

\subsection{Incommensurability in neutron scattering data}
Incommensurate peaks seen in the neutron scattering experiments of
LSCO \cite{tranquada} have been discussed as a signature of the
spin-charge fluctuations in a stripe form \cite{zaanen,salkola}.
There have been several attempts to explain the incommensurability
from the shape of the Fermi surface \cite{Si,Kao,Kuroki}.
Recently, the QP interference has been observed in the scanning
tunneling microscopy (STM) study of Bi2212
\cite{McElroyNature,Hoffman}. In another STM study, the observed
features were discussed in connection with the stripe feature
\cite{Howald}. The QP interference has been explained by the
autocorrelations of the ARPES intensity map in momentum space
\cite{McElroyPRL,Chatterjee}.

\begin{figure*}
\includegraphics[width=14cm]{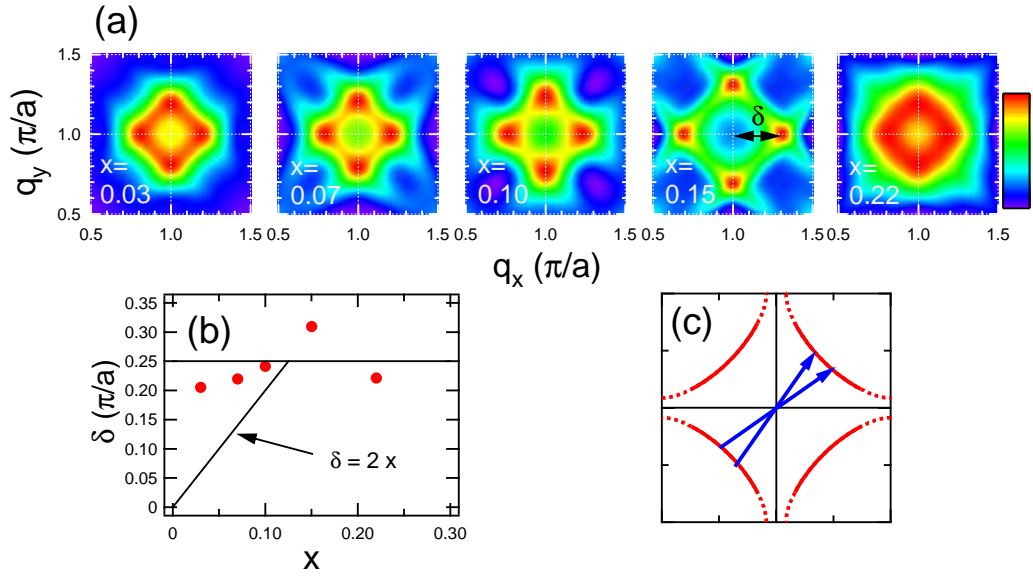}
\caption{\label{income}(Color online) Autocorrelation mapping of
the ARPES data for LSCO. (a) Intensity plots of $\Sigma_k
A(k,0)A(k+q,0)$ for various doping. (b) Doping dependence of the
incommensurability $\delta$ obtained from the autocorrelation
mapping. (c) Momentum vectors corresponding to the incommensurate
peaks for $x$=0.15.}
\end{figure*}

Therefore, we applied the autocorrelation analysis to the present
ARPES results of LSCO and found that neutron scattering peak
intensity $\rm{Im}\chi_0(q,\omega=0)$ can be approximately
expressed by the autocorrelation formula
$\rm{Im}\chi_0(q,0)\sim\Sigma_k A(k,0)A(k+q,0)$ as shown in Fig.
\ref{income}. Although this formula may not be sufficient for
correlated electron systems, autocorrelation analysis may give
some insight into the problem of the incommensurability. Figure
\ref{income} (b) shows that the obtained incommensurability
increases with hole doping, although there are quantitative
deviation from the neutron results. Also, it should be noted that
neutron scattering for $x$=0.03 shows diagonal peaks whereas our
incommensurate peaks remained vertical. In the present
autocorrelation model, such an increase of the incommensurability
can be attributed to the decreasing Fermi momentum $k_F$ measured
from the $\Gamma$ point with hole doping [Fig. \ref{kF}(b)], since
the incommensurability is given by the momentum vector connecting
nearly parallel part of the Fermi surface around the node, as
shown in the Fig. \ref{income}(c). The antinodal region does not
contribute to the autocorrelation function because spectral weight
in that region is very small due to the pseudogap opening. It is
interesting to note that the $4a\times 4a$ ordering observed in
Na-CCOC \cite{hanaguri} may be related to the nesting vector
$q=\pi/2a$ connecting the Fermi surface near ($\pi$,0)
\cite{KMShenScience}. The calculated peak width are broadened in
the overdoped $x$=0.22, which is also consistent with the neutron
results. The present analysis suggests that the shape of the Fermi
surface of LSCO with small $-t^\prime/t$ may have a nesting
instability of the observed incommensurability.  On the
quantitative level, however, the obtained incommensurability is a
little different from the stripe picture, and therefore further
analysis with more sophisticated models will be needed to address
the stripe issue.

\section{Conclusion}
We have performed a systematic ARPES study of the high-$T_c$
cuprates LSCO from the undoped insulator to the overdoped metal.
The doping dependence of the (underlying) Fermi surfaces, the QP
band dispersion and the pseudogap behaviors were studied in
detail. We observed a doping dependence of the TB parameter
$-t^\prime/t$, which is consistent with the LDA calculation, which
takes into account the effect of apical oxygen, as well as with
the CDMFT calculation, which takes into account the increased
electron correlation with decreasing hole concentration. Also, the
Fermi surface deduced using MDC's was found to approximately
fulfill Luttinger's sum rule even down to the lightly-doped region
\cite{Luttinger}.

As for the QP spectral feature, in the underdoped region, apparent
two-component behaviors occurs with the chemical potential pinning
\cite{inoCP}. The Fermi velocity was nearly doping independent
\cite{XJZhouUniversalVF} and the spectral weight of the nodal QP
increased with hole doping. The length of the Fermi arc increased
with hole doping. The relationship between the large pseudo gap
and the singularity in the DOS caused by the flat band feature is
discussed. The small pseudo gap observed around ($\pi$,0) was
found to be consistent with other experimental data.

Finally, by using the information from the ARPES results, we have
successfully reproduced the qualitative behavior of the
thermodynamic and transport properties of LSCO. In the analysis,
we demonstrated that the pseudogap and the Fermi arc feature
observed by ARPES can explain the unconventional doping dependence
of the thermodynamic and transport properties of LSCO.

Although the analysis using the ARPES data can qualitatively
explain many physical properties of the underdoped LSCO,
quantitative discrepancies remain to be explained. Obviously the
definition of the Fermi arc and its length is rather loose and
only approximate at the present stage, and will have to be further
examined on a firm ground. Also, the origin of the difference
between of LSCO and other cuprate families such as Na-CCOC remains
to be understood. The effect of different electron-phonon coupling
strengths and the effect of apical oxygen leading to the different
$t^\prime/t$ and $t^{\prime\prime}/t$ values have to be
investigated in future studies.

\section*{Acknowledgement}
We are grateful to P. Prelovsek, N. Nagaosa, A. S. Mishchenko, H.
Fukuyama, Y. Yanase, M. Ogata, O. K. Andersen, I. Dasgupta, G.
Kotliar, D. N. Basov, C. M. Ho, T. K. Lee, A. Bansil, M. Randeria
and M. Ido for enlightening discussions. This work was supported
by a Grant-in-Aid for Scientific Research in Priority Area
``Invention of Anomalous Quantum Materials", a Grant-in-Aid for
Young Scientists from the Ministry of Education, Science, Culture,
Sports and Technology, a US-Japan Joint Research Program from
JSPS, and the U.S.D.O.E. Experimental data were recorded at ALS
and SSRL which are operated by the Department of Energy's Office
of Basic Energy Science.

\end{document}